\documentclass{mem}
\usepackage{natbib}\usepackage{txfonts}\usepackage{balance}
\usepackage{graphicx}
\usepackage[a4paper,breaklinks,dvipdfm]{hyperref}
\idline{1}{1}
\begin{document}
\def\teff{$T\rm_{eff }$}
\def\kms{$\mathrm {km s}^{-1}$}
\def\lsim{\lower.5ex\hbox{$\; \buildrel < \over \sim \;$}}
\def\gsim{\lower.5ex\hbox{$\; \buildrel > \over \sim \;$}}

\title{
Nuclear interactions of low-energy cosmic rays with the interstellar medium
}

   \subtitle{}

\author{
V. \,Tatischeff
\and J. \,Kiener
          }

  \offprints{V. Tatischeff}

\institute{
Centre de Spectrom\'etrie Nucl\'eaire et de
Spectrom\'etrie de Masse, IN2P3/CNRS and Univ Paris-Sud, 91405 Orsay, France. 
\email{Vincent.Tatischeff@csnsm.in2p3.fr}
}

\authorrunning{Tatischeff \and Kiener}

\titlerunning{Nuclear interactions of low-energy cosmic-rays with the ISM}

\abstract{
Cosmic rays of kinetic energies below  $\sim$1~GeV~nucleon$^{-1}$ are thought to play a key role in the chemistry and dynamics of the interstellar medium. They are also thought to be responsible for nucleosynthesis of the light elements Li, Be, and B. However, very little is known about the flux and composition of low-energy cosmic rays since the solar modulation effect makes impossible a direct detection of these particles near Earth. We first discuss the information that the light elements have brought to cosmic-ray studies. We then discuss the prospects for detection of nuclear gamma-ray line emission produced by interaction of low-energy cosmic rays with interstellar nuclei. 
}
\maketitle{}

\section{Introduction}

Low-energy cosmic rays (LECRs) of kinetic energies $\lsim$1~GeV~nucleon$^{-1}$ are thought to be a major player in the process of star formation. They are a primary source of ionization of heavily shielded, dense molecular clouds and the resulting ionization fraction conditions both the chemistry in these regions and the coupling of the gas with the ambient magnetic field. LECRs also represent an important source of heating that contribute to hold molecular cores in equilibrium against gravitational forces. In addition, LECRs are thought to drive large-scale magnetohydrodynamic turbulence and cause amplification of magnetic field in the interstellar medium (ISM; see, e.g., Hanasz et al. in these proceedings). 

Despite LECRs being a fundamental component of the Galactic ecosystem, their composition and flux are very uncertain. This is partly because these low-energy particles cannot be detected near Earth due to the solar modulation effect. Although the Voyager 1 and 2 spacecrafts are approaching the heliopause (i.e. the contact discontinuity between the solar wind and the ISM) the twin probes will probably not be able to observe the local interstellar spectrum of CRs below 1 GeV~nucleon$^{-1}$, because of the additional modulation likely occuring in the outer heliosheath \citep[][and Fichtner in these proceedings]{sch11}. 

Thus, for still a long time our knowledge of the properties of LECRs will be based mainly on analysis of radiation produced by interaction of these particles with the ISM. In particular, a very promising way to study hadronic CRs (by far the most abundant in the Galaxy) below the kinetic energy threshold for production of neutral pions (280~MeV for $p+p$ collisions) would be to detect characteristic $\gamma$-ray lines produced by nuclear collisions of CRs with interstellar matter (Sect.~3). However, before such a detection will be made possible by a significant improvement of the sensitivity of the $\gamma$-ray space instruments, useful information on LECRs can already be obtained from measurements of cosmic abundances of the light elements Li, Be, and B (Sect.~2). 

\section{Nucleosynthesis of the light elements and origin of cosmic rays}

\begin{figure}[]
\resizebox{\hsize}{!}{\includegraphics[clip=true]{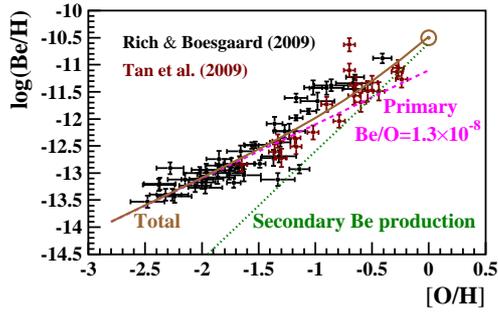}}
\caption{\footnotesize
Be abundances vs. [O/H] ([O/H]=$\log[{\rm (O/H)/(O/H)}_\odot]$, where (O/H)/(O/H)$_\odot$ is 
the O abundance relative to its solar value). Data: \citet{ric09,tan09}. The solid line shows a fit to the data with a secondary Be production component plus a primary component of constant Be/O = $1.3 \times 10^{-8}$.
}
\label{fig1}
\end{figure}

There are two basic reasons to assume that Li, Be, and B are significantly produced by nuclear interaction of galactic cosmic rays (GCRs) with the ISM. First, the abundances of these species relative to, e.g., Si, are much higher in the GCRs than in the solar system, by factors ranging from 10$^4$ to 10$^6$. Second, as first shown by \citet{ree70}, the pre-solar abundances of these elements as measured in meteorites can be approximately reproduced by assuming that they are synthesized by fragmentation of interstellar (resp. CR) C, N and O by CR (resp. interstellar) protons and $\alpha$-particles, plus a contribution from $\alpha+\alpha$ reactions for Li production, and further assuming that the GCR flux and CNO abundances have not changed throughout the history of the Galaxy\footnote{It is now admitted, however, that only $^6$Li, $^9$Be, and $^{10}$B are pure products of GCR nucleosynthesis, $^7$Li being also significantly produced in the Big Bang and in stars and $^{11}$B by $\nu$-induced spallation in core-collapse supernovae \citep[see][]{pra10}.}. Quantitatively, \citet{ree70} found that a flux of CR protons averaged over $\sim$10~Gyr $F_p(E$$>$$30~{\rm MeV})$$\sim$15~cm$^{-2}$~s$^{-1}$ would be needed to reproduce the pre-solar abundance of Li (sic). In comparison, the various estimates of the local interstellar spectrum (LIS) proposed in the literature over the past few decades give $F_p(E>30~{\rm MeV})$ in the range $\sim$5 -- 40~cm$^{-2}$~s$^{-1}$ \citep[see][and references therein]{ind09}. 

However, a full calculation of the abundances of the light elements must take into account the CR history and chemical evolution of the Galaxy. In such a framework, it was long expected that Be and B should be produced as secondary species by fragmentation of the increasingly abundant CNO nuclei in both the GCRs and the ISM, that is that Be and B abundances measured in stellar atmospheres should vary {\it quadratically} with the star metallicity. But observations made from the 1990's reveal that Be and B abundances increase {\it linearly} with metallicity (or with [O/H], see Fig.~\ref{fig1}), at least in old halo stars of low metallicity. The observed B evolution could be accounted for by a predominant production of $^{11}$B by $\nu$ spallation of $^{12}$C in core-collapse supernovae \citep{woo90}, which is a primary production process. But the observed evolution of Be is still not well understood.

\subsection{Nucleosynthesis of primary Be in supernova remnants} 

Synthesis of {\it primary} Be is expected to occur in every supernova remnant (SNR) producing CRs, because before being released in the ISM at the end of the Sedov-Taylor phase, the particles accelerated at the blast wave can interact within the remnant with SN ejecta enriched in freshly synthezised C and O nuclei  \citep{par99a,par99b}. A simple estimate of this Be production can be obtained by assuming that a constant fraction $\theta_{\rm CR}$ of the available mechanical power processed by the blast wave is continuously transformed into kinetic energy of CRs, that can produced Be while being advected with the plasma downstream the shock front. The medium inside the SNR is taken to be well-mixed and thus homogeneous. Neglecting in first approximation the adiabatic and Coulomb energy losses of the fast particles, the Be production rate per O-atom in the downstream plasma can then be estimated to be $(d{\rm Be}/dt)/{\rm O}=q_{\rm Be} \epsilon_{\rm CR}$, where $q_{\rm Be} \approx 10^{-13}$~s$^{-1}$~(erg~cm$^{-3}$)$^{-1}$~(O-atom)$^{-1}$ is the Be production rate normalized to the CR energy density, which post-shock value is $\epsilon_{\rm CR}=k\theta_{\rm CR}\rho_{\rm ISM}V_s^2$. Here, $V_s$ is the forward shock velocity, $\rho_{\rm ISM}$ the mass density of the ISM surrounding the SNR and $k$ a factor of order unity that depends on the equation of state of the shocked gas. We calculated $q_{\rm Be}$ assuming for the CR momentum spectrum a phase-space distribution resulting from diffusive shock acceleration, $f_{\rm CR}(p) \propto p^{-s}$ with $4 < s < 4.5$, and for the relative abundances of the CNO elements in the SN ejecta C:N:O=0.2:0.05:1 \citep{woo95}. 

\begin{figure}[]
\resizebox{\hsize}{!}{\includegraphics[clip=true]{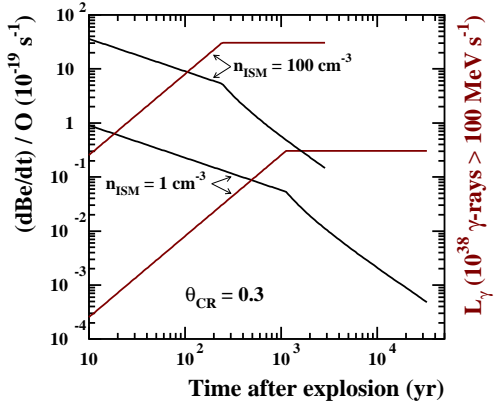}}
\caption{\footnotesize
Evolution as a function of age of the Be production rate per synthesized 0 atom ({\it black curves, left axis}) and the total luminosity in $\gamma$-rays $> 100$~MeV ({\it red curves, right axis}), for a SNR expanding into a medium of H density $n_{\rm ISM}$ (lower curves: $n_{\rm ISM}$=1~cm$^{-3}$; upper curves: $n_{\rm ISM}$=100~cm$^{-3}$).  
}
\label{fig2}
\end{figure}

Calculated Be production rates are shown in Fig.~\ref{fig2} as a function of time after the SN explosion. The evolution of $V_s$ is taken from the blast wave solution of \citet{tru99}, assuming for the supernova energy, the mass of the ejecta, and the power-law index of the outer ejecta density profile: $E_{\rm SN}=1.5 \times 10^{51}$~erg, $M_{\rm ej}=10~M_\odot$ and $n=10$, respectively. Also shown in Fig.~\ref{fig2} is the total luminosity of the SNR in $\gamma$-rays $> 100$~MeV, which we calculated from the same simple formalism but using the $\gamma$-ray emissivity $q_\gamma = 5 \times 10^{-14}$~s$^{-1}$~(erg~cm$^{-3}$)$^{-1}$~(H-atom)$^{-1}$ \citep{dru94}. We see that both the instantaneous Be production rate and the $\gamma$-ray luminosity increase with the density of the ambient ISM, but also that the blast wave evolution is more rapid for larger values of $n_{\rm ISM}$ (e.g. the end of the Sedov phase occurs at $t_{\rm rad}=31900$ and 2790~yrs for $n_{\rm ISM}=1$ and 100~cm$^{-3}$, respectively). As a result, the final Be/O abundance ratio obtained by integration of the Be production rate over the SNR lifetime obeys the following scaling relation: 
\begin{equation}
{{\rm Be} \over {\rm O}} \approx 7.1 \times 10^{-10} \bigg({\theta_{\rm CR} \over 0.3}\bigg) \bigg({n_{\rm ISM} \over 1~{\rm cm}^{-3}}\bigg)^{0.64}~. 
\end{equation}
Thus, a large ambient medium density $n_{\rm ISM} \approx 100$~cm$^{-3}$ would be needed to account for the observed abundance ratio in low-metallicity stars, $({\rm Be}/{\rm O})_{\rm obs} \approx 1.3 \times 10^{-8}$ (see Fig.~\ref{fig1}), which is not realistic. This robust conclusion is in good agreement with the results of the more detailed calculations of \citet{par99a,par99b}. Moreover, as already pointed out by \citet{fel94}, a present-day production of Be in SNRs at the level of $({\rm Be}/{\rm O})_{\rm obs}$ would be accompanied by fluxes of $\gamma$-rays $> 100$~MeV above those observed from several SNRs and the Galaxy. 

\subsection{On the origin of cosmic rays}

Thus, the observed quasi-linear increase of Be abundance with metallicity implies that a significant amount of CNO nuclei in the GCRs were -- and possibly still are -- accelerated out of freshly synthesized matter. A first model discussed in the literature is that the GCRs are accelerated mostly out of SN ejecta rather than the average ISM. Thus, \citet{lin98} suggested that high-velocity grains formed in SN ejecta could be the source of CR metals, which may be accelerated either in the forward or reverse shocks of SNe. But the absence of radioactive $^{59}$Ni in the GCRs makes this scenario unlikely, because this observation suggests a delay $\gsim$10$^5$~years between nucleosynthesis of iron-peak nuclei and their acceleration to CR energies \citep{wie99}. Indeed $^{59}$Ni decays by electron capture with a half life of $7.6 \times 10^4$~years for the neutral atoms, but this decay is suppressed once the $^{59}$Ni ions are accelerated and fully stripped. 

Another scenario is that the CNO isotopes of the GCRs are mainly accelerated from the fresh wind material of massive stars, especially Wolf-Rayet stars, as it is hit by the blast wave of the subsequent SN explosions. \citet{pra10} pointed out that models of fast rotating massive stars predict an ejection of winds enriched in CNO elements even at low metallicity and suggested that these wind sources could produce essentially constant CNO abundances in the GCRs at all times. According to \citet{ram97}, the measured Be abundances can be explained if each Galactic SN supplies on average a total energy of $\sim$10$^{50}$~erg to GCRs of current epoch composition. Assuming a typical CR acceleration efficiency of $\sim$10\%, it means that the SN blast wave has to process a large fraction of the total explosion energy ($\sim$10$^{51}$~erg) while interacting with a medium enriched in freshly synthesized CNO nuclei. In other words, the energetics of Be production implies that, on average per Galactic SN, the mass of the CNO-rich material ahead of the forward shock cannot be much lower than that of the SN ejecta. Models for non-rotating massive stars do not predict a significant ejection of CNO-rich winds at low metallicity. Those for fast-rotating, metal-poor, massive stars do display a substantial wind enriched in He-burning products after the main sequence, but only for stars of inital mass $\geq 60$~$M_\odot$ \citep{dec07}. For the Salpeter's initial mass function, such very massive stars represent no more than $\sim$5\% of all stars giving rise to SNe. We also note that the SN explosion of a fast spinning star could favor matter ejection through jets aligned along the rotation axis \citep{dec07}, such that the resulting blast wave may not interact with all the wind material. For these reasons, it is not clear to us that the scenario recently put forward by \citet{pra10} can account for enough Be production in the early Galaxy. 

A third possibility is that the GCRs are accelerated primarily out of material inside superbubbles of hot gas generated by the winds and SN explosions of massive stars formed in OB associations \citep{hig98,par99c}. The majority of core-collapse SNe likely occur in these cavities. Shock waves propagating in superbubbles can accelerate matter enriched in winds and ejecta of SN explosions of massive stars born in the same OB association, thus producing GCRs of essentially constant metallicity throughout the age of the Galaxy. This model was shown to be compatible with the absence of $^{59}$Ni in the GCRs \citep{bin08}. However, the superbubble scenario for the origin of the GCRs may suffer from two problems. First, contrary to previous results \citep[e.g.,][]{hig03}, Prantzos shows in these proceedings that the composition of superbubble gas can probably not account for the well-known anomalous $^{22}$Ne/$^{20}$Ne abundance ratio in the CR source, which is 5.3 times higher than in the solar system. Moreover, specific CR acceleration effects are expected in superbubbles due to the presence of multiple shocks and plasma turbulence between the shocks \citep{byk92,par04,fer10}. As a result, the phase-space spectrum of CRs produced in superbubbles is predicted to be harder (resp. softer) than $p^{-4}$ (the "universal" distribution produced by diffuse strong shock acceleration in test-particle regime) below (resp. above) a critical momentum that depends on the superbubble parameters \citep{fer10}. This prediction is not supported by recent {\it Fermi} LAT $\gamma$-ray observations \citep[see][]{miz11} that show the spectrum of the $\gamma$-ray-producing CRs to be spatially uniform in the outer Galaxy and consistent with the LIS (see also Tibaldo et al. in these proceedings). 

It is possible that the bulk of primary Be is in fact not synthesized by the GCRs themselves, but by a separate LECR component \citep{cas95} that could be produced in superbubbles \citep{van98}. These LECRs might be accelerated at early times during the superbubble lifetime, when the dynamics of the system is powered by the strong winds of massive stars. The standard GCRs would then be produced later on, by regular Fermi (i.e. diffusive shock) acceleration in SN blast waves first propagating in the winds of the massive progenitor stars and then in the local ISM (in order to explain the $^{22}$Ne/$^{20}$Ne abundance ratio in the GCRs, see Prantzos in these proceedings). The postulated LECR component could account for the large ionization rate deduced from H$^+_3$  abundances in diffuse interstellar clouds \citep{ind09}, although the extra ionization could also be due to LECR electrons \citep{pad09}.

\section{Nuclear gamma-ray line emission}

The most compelling evidence for a significant hadronic LECR component would come from the detection of nuclear $\gamma$-ray lines produced by interaction of these fast particles with interstellar matter. The strongest lines from the most abundant nuclei are preferentially produced in the CR energy range below a few hundred MeV per nucleon. We estimate here the diffuse Galactic emission in these lines; potential $\gamma$-ray line sources such as SNRs will be discussed elsewhere. 

\cite{ind09} proposed several LECR spectra which, added to the standard CR spectrum, would produce the observed mean ionization rate of diffuse molecular clouds $\zeta_{\rm CR} = 4 \times 10^{-16}$~s$^{-1}$. We used a slightly modified version of the broken power-law spectrum of these authors, with the CR composition measured at $\sim$1 GeV~nucleon$^{-1}$ kept constant down to 2 MeV~nucleon$^{-1}$, contrary to \cite{ind09} who used solar abundances below 200 MeV~nucleon$^{-1}$. This required a slightly changed power-law index of 1.65 for the LECR component to get the same $\zeta_{\rm CR}$. It was verified that the $\gamma$-ray emission from $\pi^0$ decay for such an interstellar CR spectrum reproduces the local $\gamma$-ray emissivity observed by {\it Fermi} LAT in the range 0.1 -- 10 GeV \citep{abd09}. 

\begin{figure}[]
\resizebox{\hsize}{!}{\includegraphics[clip=true]{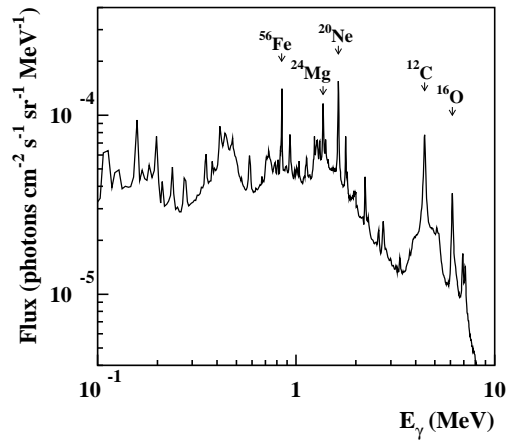}}
\caption{\footnotesize Calculated inner Galaxy nuclear $\gamma$-ray line spectrum, with indications of the emitting nuclei for some prominent lines.    
}
\label{fig3}
\end{figure}

For the calculation of the nuclear $\gamma$-ray line emission, cross sections for strong lines were taken mainly from \cite{mur09} with some modifications taking into account new nuclear data of \cite{ben11}. The production of weaker lines and $\gamma$-ray emitting radioactive nuclei was calculated with the nuclear reaction code TALYS \citep{kon08}. For the interstellar matter, we used solar abundances with 2 times solar metallicity to take into account the metal enrichment of the inner Galaxy. Parallely to the nuclear lines, the $\gamma$-ray emission following $\pi^0$ production and decay was calculated with the model of \cite{kam06}. Adjusting the latter to the pion-decay spectrum of the inner Galaxy at 0.1 -- 100~GeV as deduced from {\it Fermi} LAT observations \citep{str11} provided the absolute normalization. The calculated $\gamma$-ray line spectrum is shown in Fig. \ref{fig3}. Besides strong narrow lines, the nuclear line emission is composed of broad lines produced by the CR heavy-ion component and of thousands of weaker lines. Adding all up results in a complex and extremely rich emission spectrum. 

A very promising feature for an observation with future space-borne $\gamma$-ray telescopes is the characteristic bump above $E_\gamma \sim 3$~MeV with several strong lines of $^{12}$C and $^{16}$O. The calculated flux in the 3 -- 10 MeV band integrated over the inner Galaxy ($300^\circ \leq l \leq 60^\circ$, $\mid$$b$$\mid$ $\leq 10^\circ$) amounts to $7 \times 10^{-5}$~cm$^{-2}$~s$^{-1}$. This can be compared with the estimated continuum sensitivities of some recent $\gamma$-ray telescope projects, like, e.g., the GRIPS, DUAL and CAPSiTT proposals for the latest (2010) ESA Cosmic Vision 2015--2025 call for a Medium-size space mission. The proposed instruments feature a factor of typically 10 -- 100 sensitivity improvement with respect to {\it CGRO}/COMPTEL and {\it INTEGRAL}/SPI. Thus, the CAPSiTT 3$\sigma$ continuum sensitivity at 5 MeV in a 5-year all-sky survey for such a spatially extended emission is estimated to be $\sim 10^{-5}$~cm$^{-2}$~s$^{-1}$, which would assure a comfortable detection of the predicted nuclear line emission.  

\section{Conclusions}

The observed primary evolution of Be versus [Fe/H] (or [O/H]) shows that some CRs are accelerated out of freshly synthesized matter. The most promising sites of acceleration of these CRs appears to be superbubbles, which, however, are probably not the source of the standard GCRs detected near Earth. But superbubbles may produce a distinct component of LECRs during their early, stellar-wind-dominated phase. This supplementary CR component could account for the large ionization rate of diffuse interstellar clouds. 

Nuclear interaction $\gamma$-ray lines could provide one of the best ways of studying the origin of LECRs and their role in the Galactic ecosystem. The detection of these lines should be a major objective for the next-generation of MeV $\gamma$-ray telescopes. 
 
%\begin{acknowledgements}
%V.T. would like to acknowledge useful discussions with Nikos Prantzos. 
%\end{acknowledgements}

\bibliographystyle{aa}

\end{document}